\documentclass[doublecol,linenumbers]{epl2} 
\usepackage{amssymb}
\usepackage{amsmath}
\usepackage{subfigure}

\title{Probing linear and nonlinear microrheology of viscoelastic fluids}
\shorttitle{Probing linear and nonlinear microrheology of viscoelastic fluids} 

\author{J. R. Gomez-Solano\inst{1,2} \and C. Bechinger\inst{1,2}}
\shortauthor{Juan Ruben Gomez-Solano \etal}

\institute{                    
  \inst{1} 2. Physikalisches Institut, Universit\"at Stuttgart, Pfaffenwaldring 57, 70569 Stuttgart, Germany\\
  \inst{2} Max-Planck-Institute for Intelligent Systems, Heisenbergstrasse 3, 70569 Stuttgart, Germany
}

\abstract{
Bulk rheological properties of viscoelastic fluids have been extensively studied in macroscopic shearing geometries. However, little is known when an active microscopic probe is used to locally perturb them far from the linear response regime. Using a colloidal particle dragged periodically by scanning optical tweezers through a viscoelastic fluid, we investigate both, its linear and nonlinear microrheological response. With increasing particle velocity, we observe a transition from constant viscosity to a shear-thinning regime, where the drag force on the probe becomes a nonlinear function of the particle velocity. We demonstrate that this transition is only determined by the ratio of the  fluid's equilibrium relaxation time and the period of the driving. }

\begin{document}

\maketitle

\section{Introduction}

{Most soft materials of industrial and biological importance are viscoelastic and have non-Newtonian behavior under applied stress. For example, at sufficiently high strain rates, the deformation of their microscopic structure gives rise to nonlinear rheological response, \textit{e.g}. shear thinning or shear thickening \cite{larson}. The flow properties of such materials are usually investigated by means of controlled stress or strain rheometers in macroscopic shear geometries. These techniques provide bulk quantities (\textit{e.g.} complex shear moduli and viscosities) averaged over the entire volume of the sample, typically millilitres. However, one is often interested in local instead of bulk rheological properties of micron-sized flows, \textit{e.g.} in new synthesised materials and biological fluids, for which conventional rheological techniques are inapplicable. In such cases, colloidal probe-based techniques (microrheology) are more adequate to study flow properties in a non-invasive manner. One example is passive microrheology,  where the complex shear modulus of viscoelastic materials can be determined from the mean square displacement of embedded colloidal particles using a Generalized Stokes-Einstein relation (GSER) \cite{mason}. Due to its simple implementation and straightforward interpretation, passive microrheology is nowadays a standard technique to investigate linear rheological properties of microlitre samples of soft matter in thermal equilibrium.

Less well understood is active microrheology, where a microscopic probe is driven through the sample by an external field, \textit{e.g.} by optical tweezers, in order to directly measure the local rheological response of the fluid \cite{wilson}. Unlike bulk rheology, where the entire sample is uniformily sheared, in active microrheology strain is only applied to a small fluid volume around the probe. Therefore, instead of shear rate-stress flow curves, one usually determines velocity-force relations of the probe \cite{puertas}, which comprise memory effects of the surrounding viscoelastic fluid.
Active microrheology has been exploited to measure linear response properties in out-of-equilibrium matter, where the GSER breaks down and passive microrheology is not applicable \cite{habdas,mizuno,wilhelm,choi,gomez,bohec}. More recently, it has been proposed that active probes can be also utilized to create sufficiently strong strain and stress in complex fluids, thus providing a way to investigate nonlinear rheology at mesoscopic scales \cite{khair}. Indeed, nonlinear microrheological behavior has been induced by moving microscopic probes through dense colloidal suspensions \cite{meyer, sriram} micellar fluids \cite{cappallo,khan}, gels \cite{wilking}, and polymer solutions  \cite{gutsche,chapman}. However, it is not obvious to what extent the threshold for the nonlinear behavior is determined by the specific flow geometry of the probe rather than the properties of the investigated fluid \cite{depuit}. This is important \textit{e.g.} to understand flow properties of complex fluids in confined microscopic geometries such as in microfluidic devices \cite{perrin} and porous media \cite{scholz}, or  the motion of microswimmers in viscoelastic media \cite{arratia,lauga}.}

In this Letter, we investigate local linear and nonlinear microrheological response of a wormlike micellar solution by means of an active probe. 
Since the bulk rheology of this viscoelastic fluid under simple shear has been extensively studied \cite{lerouge}, we use it as a model system which allows a direct comparison to our active microrheological approach. For this purpose, we drag a colloidal particle along circular paths of a few microns in diameter through the solution. The particle motion creates local deformation in the fluid, whose mean rate can be varied by tuning the velocity of the probe. While at small particle velocities, the fluid can be characterized by a constant viscosity, above a certain value, we observe nonlinear behavior where its viscosity decreases dramatically with increasing particle velocity. Although this microrheological thinning effect is qualitatively similar to bulk shear thinning, we find quantitative differences in the onset of such nonlinear behavior, which depend on the size of the microrheological flow geometry. We find that, the onset of nonlinear microrheology is determined by the ratio between the local fluid's relaxation time  and the time-scale of the driving.

\begin{figure}
	\includegraphics[width=\columnwidth]{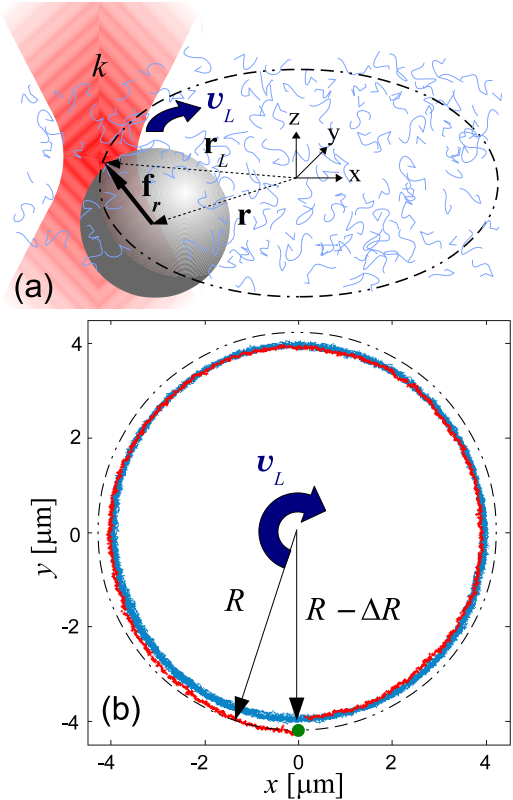}
	\caption{(Colour online)
(a) Sketch of the experimental setup for active microrheology in a viscoelastic fluid. (b) Trajectory of the probe embedded in the fluid ($c = 0.25$~wt\%, $T=35^{\circ}$C) in response to the trap's motion (dash-dotted line), at different times: $t = 0\,\mathrm{s}$ (green bullet), $0\,\mathrm{s} \le t \le 20\,\mathrm{s}$ (red dots), $20 \,\mathrm{s}< t \le 300\,\mathrm{s}$ (blue line). See text for explanation.}
\label{fig:fig1}
\end{figure}

\section{Experimental description}
In our study, we use an equimolar aqueous solution of the surfactant cetylpyridinium chloride and sodium salicylate. At surfactant concentrations $c$ above the critical micelle concentration (0.15~wt\%), wormlike micelles are formed, which give rise to linear viscoelasticity with a single stress-relaxation time $\tau$ \cite{lerouge}.  This originates from the fast breaking/recombination time of the micelles $\tau_b$ and a slower reptation time $\tau_r$:  $\tau\sim \sqrt{\tau_b \tau_r}$ \cite{berret}. In practice, we tune the value of $\tau$  by varying the surfactant concentration $c$ and the temperature $T$ of the solution, which allows us to investigate the microrheology of a viscoelastic fluid with different degrees of elasticity.
{We measure the macroscopic flow curves of these samples using a HAAKE RheoStress 1 rotational rheometer with a double-gap cylinder geometry (gaps 0.25 mm and 0.30 mm) at controlled shear rate, so that  a quantitative comparision between bulk rheology and microrheology can be done.}
In order to perform active microrheology, we add a small amount of spherical silica particles (radius $\sigma = 1.8\,\mu$m) to the solution and keep it in a thin cell (thickness $70\,\mu\mathrm{m}$), whose temperature $T$ can be adjusted between $20\pm0.1^{\circ}$C and $40\pm 0.1^{\circ}$C by a flow thermostat. A single particle is dragged through the liquid by a scanning optical tweezers, which is created by deflection of a Gaussian laser beam ($\lambda = 1070$~nm) on a galvanostatically driven pair of mirrors and subsequent focusing by a microscope objective ($100\times$, NA = 1.3) into the sample. 
The focal plane is created $20\,\mu\mathrm{m}$ apart from the lower wall to avoid hydrodynamic interactions. 
The optical trap at position ${\bf{r}}_L$ exerts a restoring force on the particle at position $\bf{r}$, \textit{i.e.} ${\bf{f}}_r=-k ({\bf{r}} - {\bf{r}}_L) $, as sketched in fig.~\ref{fig:fig1}(a). At fixed surfactant concentration $c$, temperature $T$ and constant laser power, we determine the spring constant $k$ of the trap from the variance $\langle \delta x^2 \rangle$ of the particle's positional fluctuations at thermal equilibrium by means of the equipartition theorem $ k = {k_B T}/{\langle \delta x^2 \rangle}$. The value of $k$ was adjusted between  $1.5\,\mathrm{pN}\,\mu\mathrm{m}^{-1}$ and $12.0 \,\mathrm{pN}\,\mu\mathrm{m}^{-1}$ by tuning the laser power between 40~mW and 200~mW.
In order to drag the particle through the fluid, we move the trap at constant speed $v_L$ along circular paths of radius $R$, ${\bf{r}}_L(t) = \left(-R \sin \frac{v_L t}{R},  -R \cos \frac{v_L t}{R}\right)$, in such a way that the particle moves at the same angular velocity as the trap. The value of $R$ was tuned between  $2.10\, \mu$m and  $5.25\, \mu$m to investigate the possible dependence of the microrheological response of the fluid on the size of the flow geometry.  Note that this geometry guarantees that every portion of the fluid on the circular path is revisited at every revolution of the particle, thus directly probing its local relaxation time.

{When moving at constant velocity $v$, the particle deforms the surrounding fluid, creating a combination of shear, compression and extensional flow \cite{depuit1}. Then, in order to compare the microrheological flow curves to those obtained by bulk rheology, the characteristic microrheological shear rate must be estimated. Since the maximum Reynolds number is $\mathrm{Re} \sim O(10^{-4})$, we approximate the velocity profile of the fluid around the particle to that of a Stokes flow, which yields a local shear rate $\dot{\gamma} = 3 \pi v / (8\sigma)$ spatially averaged over the particle's surface. }

In fig.~\ref{fig:fig1}(b) we show a typical trajectory ${\bf{r}}(t)$ of the probe in response to the motion of the trap.
The starting point ${\bf{r}}(0)$, which defines the time $t=0 \,\mathrm{s}$, is shown as a green bullet and coincides with the initial position of the trap. Then, at $t  > 0 \,\mathrm{s}$, the trap begins to move at speed $v_L$ giving rise to a transient of the particle's trajectory ($0\,\mathrm{s} < t \le 20\,\mathrm{s}$, red dots) during which its radial position changes until it reaches a steady state. In a steady state ($20 \,\mathrm{s}< t$, blue line), the probe describes a circular trajectory at constant velocity. Although the particle moves at the same angular velocity as the trap, the particle's position lags not only axially but also radially a distance $\Delta R$ behind the centre of the trap due to the drag force exerted by the surrounding fluid, as shown in fig.~\ref{fig:fig1}(b). Therefore, the radial position of the particle, its velocity and its position relative to the centre of the trap are  $R - \Delta R < R$, $v = (1 - \Delta R/R)v_L < v_L $ and $|{\bf{r}} - {\bf{r}}_L| =  R[1 - (1-\Delta R/R)^2]^{1/2}$, respectively. In this steady state,
the fluid resistance can be characterized by an effective viscosity $\eta$, such that the drag force on the particle can be expressed as $f_d = 6\pi \sigma \eta v$. Consequently, the balance between the viscous drag and the restoring force, $ f_d = f_r$  can be written as
\begin{equation}\label{eq:eq1}
	 6\pi \sigma \eta v_L \left(  1 - \frac{\Delta R}{R} \right) = kR\sqrt{1-\left( 1 - \frac{\Delta R}{R} \right)^2}.
\end{equation}
Eq.~(\ref{eq:eq1}) allows us to determine the local viscosity of the fluid $\eta$ and the drag force $f_d$ at a given mean shear rate $\dot{\gamma}$ by measuring $\Delta R$ from steady-state particle's trajectories.

\begin{figure}
	\includegraphics[width=\columnwidth]{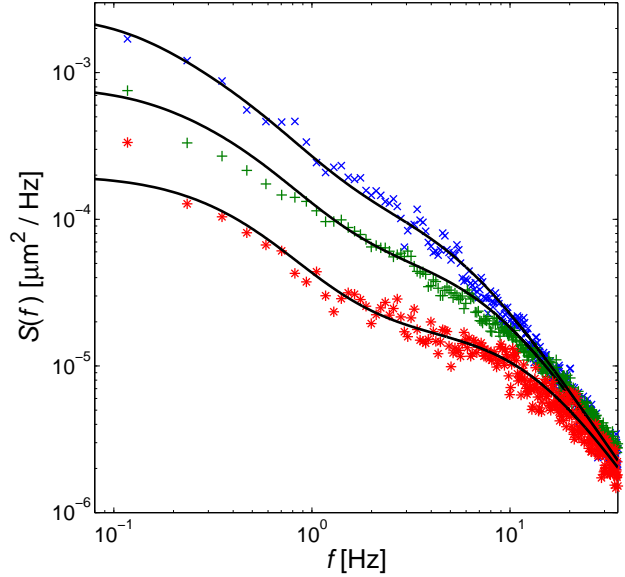}
	\caption{(Colour online)
Power spectral density of the positional fluctuations of the particle trapped by the twezers at different values of $k$ in the solution at $c = 0.18$~wt\%  and $T = 37^{\circ}$C:  $k = 2.5\times 10^{-6}$N/m ($\times$), $k = 5.3\times 10^{-6}$N/m ($+$), and $k = 1.1\times 10^{-5}$N/m ($\ast$). The solid lines are fits to eqs.~(\ref{eq:eq2}) and (\ref{eq:eq3}) with fitting parameters $\eta_0 = 0.043 \pm 0.004$~Pa~s, $\eta_\infty = 0.005 \pm 0.001$~Pa~s, and $\tau = 0.6 \pm 0.1$~s for the three different values of $k$.}
\label{fig:fig2}
\end{figure}

A prior analysis of the equilibrium dynamics of the particle trapped by the optical tweezers at rest ($v_L = 0$) is crucial to determine the local linear rheological properties of the fluid by passive microrheology \cite{buchanan}. In particular, the stress relaxation time $\tau$, which quantifies the memory effects of the fluid due to the elasticity of the wormlike micelles, can be computed from the power spectral density $S$ of the particle's positional fluctuations. For example, in fig.~\ref{fig:fig2} we plot $S$ as a function of the frequency $f$ for different values of $k$ at  $c = 0.18$~wt\% and $T = 37^{\circ}$C. Since at $v_L = 0$ the system is in thermal equilibrium and the micelles are much smaller than the probe's size, the GSER relation allows to write $S(f)$ in terms of the complex shear modulus of the fluid $G^*(f) = G'(f) + \mathrm{i}G''(f)$ \cite{mizuno2}
\begin{equation}\label{eq:eq2}
S(f) = \frac{2 k_B T}{\pi f} \frac{6 \pi \sigma G''(f)}{[k + 6\pi \sigma G'(f)]^2 + [6\pi \sigma G''(f)]^2}.
\end{equation}
In order to compute $\tau$ from the measured power spectral densities, in eq.~(\ref{eq:eq2}) we use the Jeffreys model for $G^*$. This linear rheological model incorporates two dissipative mechanisms associated to the zero-shear viscosity $\eta_0$ and the solvent viscosity $\eta_{\infty}$, with a single elastic element with shear modulus $G_{\infty} = (\eta_0 - \eta_{\infty})/\tau$, which accounts for the energy storage \cite{raikher}. The corresponding storage $G'$ and loss $G''$ moduli are given by
\begin{eqnarray}\label{eq:eq3}
	G'(f) & = & \frac{4\pi^2 f^2 \tau (\eta_0 - \eta_{\infty})}{1 + 4\pi^2 f^2 \tau^2},\nonumber\\
	G''(f) &=&2\pi f \eta_{\infty} + \frac{2\pi f (\eta_0 - \eta_{\infty})}{1+4 \pi^2 f^2 \tau^2},
\end{eqnarray}
respectively \cite{grimm}. We find that the experimental power spectra can be fitted to those of a particle in a Jeffreys fluid, described by eqs.~(\ref{eq:eq2}) and (\ref{eq:eq3}), as shown by the solid lines in fig~\ref{fig:fig2}. In order to perform this nonlinear fit, we use the values of $k$ obtained from previous calibration, so that the fitting parameters are $\eta_0$, $\eta_{\infty}$ and $\tau$. We observe that at constant $c$ and $T$, the values of $\eta_0$, $\eta_{\infty}$ and $\tau$ are constant within the experimental errors regardless of the value of $k$, as plotted in fig.~\ref{fig:fig2}. This method allows us to determine $\tau$ unambiguously for the different solutions investigated in the following.

\section{Results and discussion}

\begin{figure*}
	\includegraphics{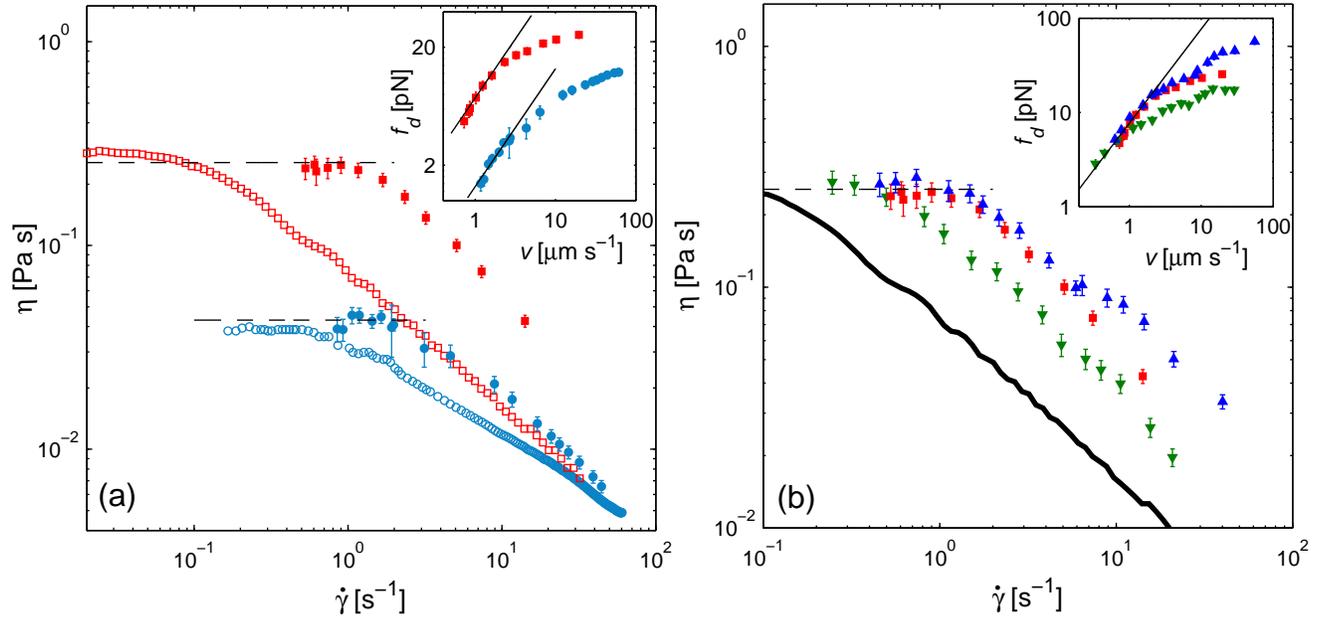}
\caption{(Colour online)
(a) Microrheological flow curves of the micellar solution  at two different concentrations and temperatures: $c = 0.18$~wt\%, $T = 37^{\circ}$C ($\bullet$) and $c = 0.25$~wt\%, $T = 30^{\circ}$C ($\blacksquare$). The corresponding zero-shear viscosities $\eta_0$ are represented as dashed lines. The open symbols are the corresponding bulk measurements. Inset: Corresponding drag force as a function of particle velocity. Same symbols as in main plot. The solid lines represent the Stokes' law. (b) Viscosity at $c = 0.25$~wt\% and $T = 30^{\circ}$C as a function of mean shear rate for different values of $R$: $2.10 \, \mu\mathrm{m}$ ($\blacktriangledown$), $4.20 \, \mu\mathrm{m}$ ($\blacksquare$), and $5.25 \, \mu\mathrm{m}$ ($\blacktriangle$). The thick solid line is the corresponding bulk measurement. Inset: Drag force as a function of particle velocity for the three different values of $R$. The solid line represents the Stokes' law.}
\label{fig:fig3}
\end{figure*}

In fig.~\ref{fig:fig3}(a) we plot as solid symbols the dependence of $\eta$ on $\dot{\gamma}$ for solutions at two different surfactant concentrations $c$ and temperatures when dragging the probe along a circle of radius $R=4.2\,\mu\mathrm{m}$. Similar microrheological flow curves are  obtained for other values of $c$ and $T$ and different radius $R$ and all of them exhibit the same features (data not shown). For comparison, we also plot as open symbols the corresponding macroscopic flow curves of micellar solutions under the same experimental conditions as in active microrheology.
We observe that at sufficiently small $\dot{\gamma}$, the viscosity is constant $\eta=\eta_0$. For example, $\eta_0 = 0.043$~Pa~s and $\eta_0 = 0.250$~Pa~s at $c = 0.18$~wt\%, $T = 37^{\circ}$C and $c = 0.25$~wt\%, $T = 30^{\circ}$C, respectively, as depicted by the dashed lines in fig.~{\ref{fig:fig3}}(a).  The existence of a well-defined zero-shear viscosity $\eta_0$ is a consequence of the Stokes's law, \textit{i.e.} the drag force on the probe is a linear function of the small particle velocity $v$, $f_d = 6\pi \sigma\eta_0 v$, as shown in the inset of fig.~\ref{fig:fig3}(a).
We find that the values of $\eta_0$ are in good agreement with those obtained by passive microrheology ($v_L = 0$), and by bulk rheology under controlled shear rate. The agreement between different methods results from the linear response of the fluid, which is a property independent of the details of the applied stress.

We observe that, when increasing further the particle velocity, the fluid exhibits shear thinning, \textit{i.e.} the viscosity decreases dramatically with increasing $\dot{\gamma}$ above certain onset. This is because the fluid structure is perturbed far away from the linear response regime by the particle motion, thus probing its non-Newtonian properties. The nonlinear rheological response of the fluid translates into a breakdown of the Stokes' law at large $v$, as shown in the inset of fig.~\ref{fig:fig3}(a).
Note that the thinning behavior of $\eta$ with increasing $\dot{\gamma}$ in the microrheological experiment is qualitatively similar to the bulk behavior plotted as open symbols in fig.~\ref{fig:fig3}(a). However, we observe quantitave differences between both kinds of measurements. 
The onset of microrheological shear thinning occurs at values of $\dot{\gamma}$ higher than those in bulk measurements. This observation suggests that, unlike bulk rheology, the onset of nonlinear microrheological behavior depends on the size of the flow geometry. Indeed, in fig.~\ref{fig:fig3}(b) we show that, even for the same fluid at fixed $c$ and $T$, \textit{i.e.} at fixed $\tau$, there is a systematic deviation
of the microrheological curves with respect to the bulk measurement with increasing radius $R$ of the circular path. 
In particular, both the onset of shear thinning is shifted to higher values of $\dot{\gamma}$ and the deviations from the Stokes' law occurs at higher values of $v$ with increasing $R$, as plotted in the inset of fig.~\ref{fig:fig3}(b).
These  results demonstrate that, in a microrheological experiment, nonlinear rheological behavior of viscoelastic fluids is not simply determined by $\dot{\gamma}$ but it dependends on the size of the flow geometry. In a bulk measurement with a rheometer, the complete sample is uniformly sheared by the confining walls. Therefore, the global time-scale imposed by the shear
is $\sim \dot{\gamma}^{-1}$ regardless of the details of the shearing
geometry. However, when using a microscopic probe, stress is applied only to a micron-sized fluid volume.
Once the probe leaves this volume, the fluid structure relaxes back to its equilibrium state. Therefore, in order
to prevent the complete relaxation of this small volume, the probe must revisit it within a time-scale comparable or smaller than $\tau$, which in our experimental configuration corresponds to the revolution period of the probe $\tau_L = 2 \pi R / v_L$.

\begin{figure}
	\includegraphics[width=\columnwidth]{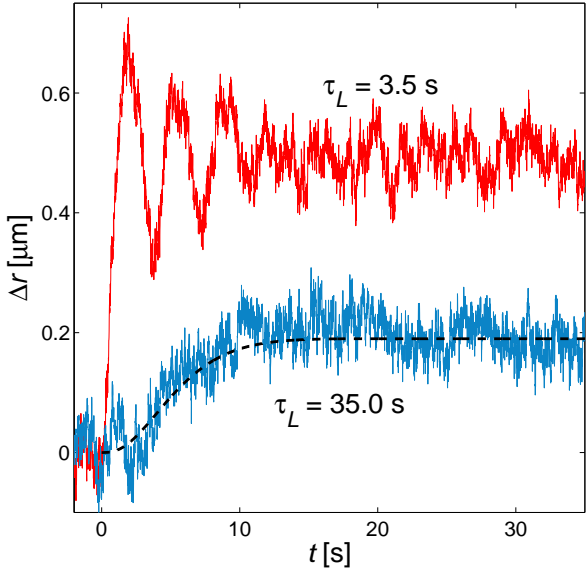}
\caption{(Colour online)
Time evolution of the radial displacement of the probe (fluid at $c = 0.25$\%wt and $T = 30^{\circ}$C) with respect to ${\bf{r}}_L$ moving along a circle of radius $R = 2.10 \, \mu\mathrm{m}$ with two  different revolution periods $\tau_L$. At time $t < 0$~s the trap is at rest, whereas at $t \ge 0\,\mathrm{s}$ it moves at constant velocity. Dashed line: time evolution of $\Delta r$ for a particle dragged by the trap with $\tau_L = 35.0\,\mathrm{s}$ in a Newtonian fluid with $\eta_0 = 0.250$~Pa~s.}
\label{fig:fig4}
\end{figure}

The developement of microrheological shear thinning can be better understood by observing the transient dynamics of the particle after starting the motion of the trap at time $t = 0$~s, like the trajectory shown as red dots in fig.~\ref{fig:fig1}(b). In fig.~\ref{fig:fig4} we plot the time evolution of the radial displacement of the particle, $\Delta r(t) \equiv R - |{\bf{r}}(t)|$, dragged by the trap at two different values of $\tau_L$ through the solution at $c = 0.25$~wt\% and $T = 30^{\circ}$C. For these parameters, the stress relaxation time of the fluid is $\tau = 2.2\pm 0.4$~s. We observe that, when moving slowly the trap with a revolution period $ \tau_L=$~35~s, $\Delta r$ saturates to a constant steady-state value (blue line). At this value of $\tau_L$, which is much larger than $\tau$, every portion of the fluid on the circle has enough time to relax the stress previously stored by the local deformation even when the particle revisits it periodically. 
This should translate into a drag force $f_d$ which obeys the Stokes' law with shear-independent viscosity $\eta_0$ and consequently, Newtonian behavior is expected to occur. Indeed, we find that in this case the time evolution of $\Delta r$  corresponds to that in a Newtonian fluid with viscosity $\eta_0 = 0.25$~Pa~s (dashed line). Therefore, we verify that only linear rheological response of the fluid is probed when $\tau_L \gg \tau$. 
More complex dynamics of $\Delta r$ are observed when $\tau_L$ is comparable or smaller that $\tau$. In fig.~{\ref{fig:fig4}} we plot in red the time evolution of $\Delta r$ when moving the trap at $\tau_L = 3.5$~s. In such a case, $\Delta r$ increases monotonically during the first 2~s, but then it decreases when the particle revisits the pre-probed fluid volume within a time $\tau_L$.
This is because the particle initially undergoes the viscous drag determined by $\eta_0$, which translates into a comparatively large $\Delta r$. Once the particle tries to build up a steady flow profile, it induces a rearrangement of the micelles, which in turn reduces the viscous drag, thus decreasing $\Delta r$. Next, $\Delta r$ exhibits damped oscillations with a period equal to $\tau_L$, reaching finally a constant steady-state value at $t > 10$~s. These oscillations reveal that the fluid attempts to recover its unperturbed structure with large $\eta$, and therefore the particle needs to perform a sequence of several revolutions in order to pre-shear the micellar solution and create a persistent state at $10\,\mathrm{s}\,\le t$ with reduced viscosity $\eta= 0.110$~Pa~s~$\,<\eta_0$. This observation suggests that a fast periodic motion is necessary to reach a nonlinear regime, inducing an orientational order of the micelles, which would otherwise have enough time to relax back to its quiescent state.
It should be noted that, unlike the nonlinear microrheology of colloidal suspensions \cite{meyer,sriram}, a linear motion of the probe is not enough to induce the strong thinning effect that we observe in the micellar solution. In fact, we check that for the same values of $k$ and $\dot{\gamma}$ used in fig.~\ref{fig:fig3}(b) for which we observe shear thinning, we
are not even able to drag continuously the particle along a linear trajectory, where $R\rightarrow \infty$ and $\tau_L  \rightarrow\infty$. This is because, in absence of pre-shear, the viscosity of the fluid around the particle relaxes to the linear-response value $\eta_0$, which gives rise to a viscous drag much larger than the maximum restoring force exerted by the trap\footnote{{See supplementary videos which illustrate either the local relaxation of the fluid structure or the development of microrheological shear thinning for a particle driven by the tweezers along different trajectories at the same $v  = 16 \, \mu \mathrm{m}\,s^{-1}$.}}.

\begin{figure}
	\includegraphics[width=\columnwidth]{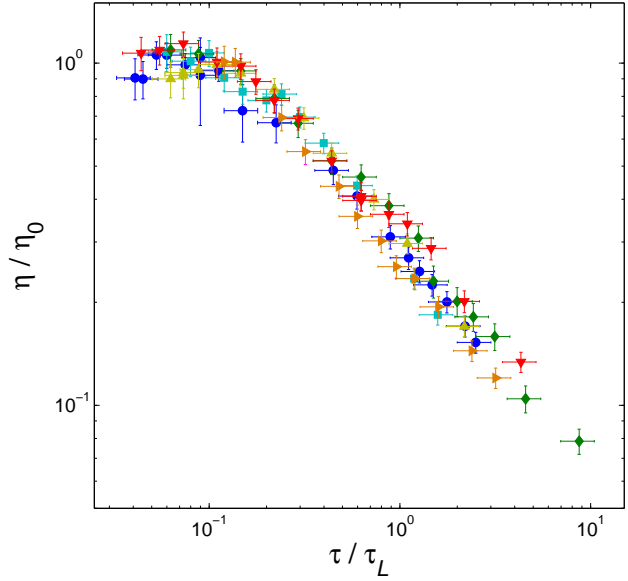}
\caption{(Colour online)
Normalized viscosity $\eta/\eta_0$ as a function of $\tau / \tau_L$ for different $(c,T)$ and $R$: $c = 0.18$~wt\%, $T = 37^{\circ}$C, $R=4.20\,\mu\mathrm{m}$ ($\bullet$), $c = 0.22$~wt\%, $T = 35^{\circ}$C, $R=4.20\,\mu\mathrm{m}$ ($\blacksquare$),  $c = 0.25$~wt\%, $T = 30^{\circ}$C, $R=2.10\,\mu\mathrm{m}$ ($\blacklozenge$), $c = 0.25$~wt\%, $T = 30^{\circ}$C, $R=4.20\,\mu\mathrm{m}$ ($\blacktriangle$), $c = 0.25$~wt\%, $T = 30^{\circ}$C, $R=5.25\,\mu\mathrm{m}$ ($\blacktriangledown$), and $c = 0.29$~wt\%, $T = 40^{\circ}$C, $R=5.25\,\mu\mathrm{m}$ ($\blacktriangleright$).
}
\label{fig:fig5}
\end{figure}

A  dimensionless quantity which characterizes the degree of nonlinearity in bulk rheology of viscoelastic fluids is the Weissenberg number, defined as the product of the rate of deformation times the relaxation time of the fluid. Thus, nonlinear bulk behavior typically occurs for $\dot{\gamma} \tau \gtrsim 1$  under simple macroscopic shear.  The previously described dependence of the onset of shear thinning on $R$ and the analysis of the transient  probe's trajectories suggest that, in active microrheology of viscoelastic fluids, the quantity $\tau / \tau_L$ plays the role of the Weissenberg number instead of $\dot{\gamma}\tau$. 
In fig.~{\ref{fig:fig5}} we plot the normalized viscosity $\eta/\eta_0$ as a function of $\tau / \tau_L$ for different values of $R$ and for different ($c,T$), which span values of $\tau$ in the interval $0.6-4$~s and $\eta_0$ in the interval $0.043-0.910$~Pa~s. Interestingly, all curves collapse onto a master curve, which confirms that unlike bulk rheology, the transition from linear to nonlinear microrheology is determined by the ratio between  $\tau$ and the time-scale of the periodic driving and not by $\dot{\gamma}\tau$. We point out that, despite the resemblance to the temperature-concentration superposition of micellar solutions under simple steady shear \cite{lerouge}, which relies on a bulk relaxation time of the entire sample, the microrheological superposition is achieved by means of a local relaxation time directly probed by tuning $\tau_L$. Depending on the value of $\tau_L$, the local relaxation of the fluid's structure or the subsequence development of shear thinning can be unequivocally inferred from the transient probe's trajectories.
Thus, our results support the idea that active microrheology is not simply a surrogate of  macrorheology but it can uncover properties of viscoelastic fluids that cannot be investigated by means of bulk measurements \cite{puertas}.

\section{Conclusion}

In conclusion, we have studied linear and nonlinear active microrheology of a wormlike micellar solution. 
Although we find qualitative similarities between micro and bulk rheology, namely the transition from linear to nonlinear microrheological behavior with increasing particle velocity, quantitative differences emerge from the microscopic flow geometry. In particular, by analysing the transient particle's dynamics after startup of the trap's motion, we observe that microrheological thinning develops from the interplay between the equilibrium stress relaxation time of the fluid and the time-scale imposed by a periodic driving. Thus, we show that the onset of nonlinear microrheological response is determined by the ratio between these two time-scales, which plays the role of a Weissenberg number.

\acknowledgments
We would like to acknowledge Christian Scholz for helpful discussions.

\end{document}